\documentclass[sigconf]{acmart}
\AtBeginDocument{%
  }

\copyrightyear{2026}
\acmYear{2026}
\setcopyright{cc}
\setcctype{by}
\acmConference[RecSys '26]{20th ACM Conference on Recommender Systems}{September 27-October 02, 2026}{Minneapolis, MN, USA}
\acmBooktitle{20th ACM Conference on Recommender Systems (RecSys '26), September 27-October 02, 2026, Minneapolis, MN, USA}
\acmDOI{10.1145/3773078.3831835}
\acmISBN{979-8-4007-2284-4/2026/09}

\usepackage{soul}
\usepackage{booktabs}
\usepackage{graphicx}
\usepackage[inline]{enumitem}
\usepackage{framed}

\begin{document}

\title{The Utility of LLMs in Recommender Systems Explanation Evaluation}

\author{Kathrin Wardatzky}
\email{wardatzky@ifi.uzh.ch}
\orcid{0000-0002-7043-7326}
\affiliation{%
  \institution{University of Zurich}
  \city{Zurich}
  \country{Switzerland}
}

\author{Oana Inel}
\email{inel@ifi.uzh.ch}
\orcid{0000-0003-4691-6586}
\affiliation{%
  \institution{University of Zurich}
  \city{Zurich}
  \country{Switzerland}
}

\author{Luca Rossetto}
\email{luca.rossetto@dcu.ie}
\orcid{0000-0002-5389-9465}
\affiliation{%
  \institution{Dublin City University}
  \city{Dublin}
  \country{Ireland}
}

\author{Abraham Bernstein}
\email{bernstein@ifi.uzh.ch}
\orcid{0000-0002-0128-4602}
\affiliation{%
  \institution{University of Zurich}
  \city{Zurich}
  \country{Switzerland}
}

\renewcommand{\shortauthors}{Wardatzky, et al.}

\raggedbottom

\begin{abstract}
Explanations play a crucial role in creating trustworthy recommender systems (RS), yet choosing a good explanation method presents challenges. 
Many explanation methods exist, but little guidance exists on which is best for which setting.
Existing explanation generation methods often produce abstract outputs that require further formatting to become user-friendly, with a seemingly endless pool of options. 
Running user-based evaluations of all possible options is usually unfeasible, while automated evaluation metrics often either assess only the explainer's abstract output or require comparison with a ground truth, which is generally unavailable. 
Recent studies have shown that large language models (LLMs) can serve as ``judges'' for explanation evaluation, but their reliability has not yet been thoroughly explored. 
This paper studies the utility of LLMs in selecting an effective explanation method for a given application. 
We first explore their ability to generate explanation prototypes given varying information about the RS and the user.
Specifically, we generate 18 distinct explanation prototypes, which are subsequently evaluated by 14 LLMs of varying sizes across two temperature settings.
We compare these against human ratings derived from a user study. Our results show that while LLMs exhibit human-like rating patterns and achieve moderate rank correlation with human raters, their absolute rating agreement is low and varies substantially by model size and evaluation construct. We derive four practical recommendations: keep explanation-generation prompts concise, prefer larger models for evaluation, pre-test evaluation constructs, and audit explanations for factual accuracy, as neither humans nor LLMs reliably detect non-factual content.
\end{abstract}



\maketitle

\section{Introduction}
Explainable AI has been a popular tool for increasing, among other factors, perceived transparency, user trust, satisfaction, and efficiency in recommender systems (RS) ~\cite{tintarev2015explaining}.
Recent years have seen an increase in the development of explanation methods for various families of RS approaches \cite{kouki2019personalized,gedikli2014should,markchom2025review,tan2021counterfactual}. Accordingly, frameworks such as recoXplainer \cite{coba2022recoxplainer} include a variety of intrinsically explainable RS and post-hoc approaches, enabling their benchmarking. 

However, evaluating RS explanations is not trivial, as it lacks agreed-upon standards and guidelines, making it difficult to determine which approach works best in a given application scenario~\cite{wardatzky2024_TORS}. On the one hand, automated evaluation metrics fail to capture explanation quality holistically and are often developed for classification problems~\cite {wardatzky2025toward}. For instance, metrics often evaluate the structured output of an explainer (e.g., Mean Explainability Precision \cite{DAMAK2022100208} or Feature Diversity \cite{Li2020_generateneural}). On the other hand, often, the output of explainers needs further formatting to make it accessible and user-friendly. 
We currently do not know much about how explanations should be designed to yield a desired effect under specific contextual variables, such as the application domain or user characteristics \cite{wardatzky2024_TORS}.
Therefore, the number of possible explanation formats that an application provider needs to consider is seemingly endless.
While metrics that evaluate formatted explanations and could be used to select appropriate explanations also exist (e.g., BLEU~\cite{BLEU}, ROUGE~\cite{lin-2004-rouge}), they often require a ground truth to compare the explanation to, which usually needs to be approximated, e.g., with user reviews, or is simply unavailable. 
User-centric evaluations provide insights into the practical utility of explanations. However, setting up user evaluations is costly and time-consuming, thereby hindering the iterative improvement of explanations.

Recently, Large Language Models (LLMs) have been increasingly used in the evaluation of various applications, such as document retrieval \cite{penha2025llm}, translation quality evaluation  \cite{kocmi2023large}, and even recommendation explanation evaluation \cite{zhang2024large, Li2025_alert, Zhao2025_adaptingLLMs}. 
They are comparatively cheap to run, enabling large-scale evaluations that would not be feasible with human users.
Related work has shown promising results of using LLMs to assess the reasoning and linguistic quality of explanations \cite{Li2025_alert}, their relevance, persuasiveness, and informativeness \cite{Zhao2025_adaptingLLMs}, or their transparency, and accuracy \cite{zhang2024large}. 

Existing literature often focuses on the final evaluation stage of explanations; this paper investigates LLMs' potential to evaluate RS explanations during prototyping.
In this stage, one encounters a multitude of possible user-friendly explanations from which one needs to select a subset for human evaluation.
We use a knowledge graph-based RS to generate recommendations and to extract the explanation, i.e., path-trace leading from the target user to the recommended item.
We then use an LLM to generate 18 different explanation types for these recommendations. 
The explanations are generated by combining four information aspects---user expertise, user history, recommendation information, and explanation goal---about the user preferences and the RS described in the prompt. 
We use this approach to simulate the explanation formatting stage, in which a decision must be made on how to format the structured output of an explainable RS to best serve the user group and application domain.
We explore the capabilities of 14 LLMs across two temperature settings to rate explanations on six quality dimensions and also compare these results with human assessments.

In contrast to existing work, our focus is not primarily on the accuracy of the model ratings but rather on the overlap of rating patterns, ranking, and rating/ranking agreement compared to human evaluators.
Finally, we evaluate a subset of the explanations in a user study to compare human and LLM rating behavior.
In summary, this paper contributes the following:\footnote{Code and data can be found here: \url{https://gitlab.ifi.uzh.ch/DDIS-Public/llm-eval-utility}.}
\begin{enumerate}[leftmargin=0.5cm]
    \item an extensive analysis of systematically LLM-generated explanations using different information aspects about the RS and user preference in the prompt,
    \item a large-scale LLM-as-judge evaluation across 14 models covering large/small open-weight models and proprietary models along six quality dimensions, analyzing the evaluator reliability, 
    \item a large-scale user study with $N=216$ participants comparing the LLM ratings with human ratings, and
    \item guidelines on how to use the LLM-as-a-judge paradigm to select an effective explanation method in a given application.
\end{enumerate}

\section{Related Work}
In this section, we discuss related work on explanation generation using LLMs, explanation evaluation, and evaluation ratings.

\subsection{LLM-based Explanation Generation in RS}
Generative AI models have been extensively integrated into the explanation generation process either to replace RS and explainers \cite{Lei2024_RecExplainer} or to transform explainers' output into user-friendly explanations \cite{luo2024unlocking}. \citet{Said2024_LLMReview} found that LLM-generated explanations provide more flexibility and personalization opportunities compared to standard explanation methods and are often favored by users. \citet{Okoso2025_tone-aware} used LLMs to generate explanations for six tones (e.g., humorous or formal) across three application domains and found that the effect of tone-adapted explanations depends on the domain and user attributes. \citet{luo2024unlocking} and \citet{Lei2024_RecExplainer} prompted LLMs with different information aspects about the user and the recommended item, such as the user's interaction history, user profile, or item features. Nevertheless, \citet{Said2024_LLMReview} highlights that LLMs tend to provide excessive detail that can overwhelm users and that there remains a lack of evaluation methods tailored to LLM explanations. Furthermore, the explanations seem to be influenced by attributes such as user profiles, item categories, and user history size \cite{Lei2024_RecExplainer}.

\subsection{Explanation Evaluation in RS}
Explanations in RS are commonly evaluated using automated offline metrics or user studies. 
Frameworks such as RecoXplainer \cite{coba2022recoxplainer} and Hopwise \cite{Boratto2025_hopwise} offer unified evaluation approaches by providing explainable RS, post-hoc explainability methods, and metrics to evaluate the precision, faithfulness, or quality attributes of the explainable outputs. However, such metrics evaluate the explainer's abstract output rather than the explanation that can be shown to a user. In addition, while already established metrics such as BLEU~\cite{BLEU} and ROUGE \cite{lin-2004-rouge} can be used to assess feature hallucination~\cite{Ariza-Casabona2024_text-based-eval}, they seem to be specific to their recommender application. Nevertheless, alignment between offline metrics and user and explanation goals is an ongoing research direction, with a recent study by \citet{Zanon2025_offlinemetrics} proposing that both online and offline evaluation are needed, but that greater emphasis should be placed on developing user-centric metrics. In the health domain, \citet{Hulstijn2023_metrics} argue that objective and subjective evaluation metrics are complementary and evaluations should use both.

User-based evaluation of RS explanations is often conducted as an online study to assess users' perceptions (for a review, see~\cite{wardatzky2024_TORS}).
Occasionally, explanations are evaluated in a running system with an A/B test (e.g., \citet{takami_personality-based_2023}) or in a controlled lab experiment using methods such as eye-tracking (e.g., \citet{Coba2019_decision}) or card-sorting (e.g., \citet{Tsai2019_evaluating}).

Recent LLM developments, though, have seen an increased adoption of the LLM-as-a-judge paradigm for explanation evaluation~\cite{zhang2024large, Li2025_alert, Zhao2025_adaptingLLMs}. \citet{Zhao2025_adaptingLLMs} combine Model-Agnostic Meta-Learning with LoRA-based parameter-efficient tuning and evaluate the agreement between the generated scores and a ground truth extracted from a simulated and a human-annotated dataset. \citet{zhang2024large} found that GPT-4 can be an accurate and cost-efficient evaluation solution compared to traditional human-based approaches. In addition, the evaluation's accuracy can be improved by aggregating multiple heterogeneous zero-shot LLMs. \citet{Li2025_alert} propose two LLM-based evaluators: a generative evaluator that performs pairwise comparisons of explanations along the defined criteria, and a discriminative evaluator that assigns a score to the generative evaluator's output. They showed a high correlation with human annotators for a fraction of the costs of human evaluation.

User evaluation is typically conducted by collecting users' opinions via rating scales. A large body of literature acknowledges that rating scales are not uniformly used by study respondents. Differences in ratings typically arise from different perceptions or meanings of rating scales, a tendency to select extreme or neutral ratings \cite{van2004response, uher2018quantitative}, or from possible biases due to priming effects, or user personality \cite{aniceto2023influence}. Some of these aspects have also been investigated for LLMs-as-a-judge approaches. Recent research has shown that LLM judges tend to favor answers from the same family of models \cite{panickssery2024llm} or answer options that are provided later in a list \cite{wang2024large}, and their answers are not always self-consistent across scales \cite{li2026grading}. 

\subsection{Our Contributions}
We generate an extensive set of user-friendly explanations using 14 LLMs, covering various families, sizes, and reasoning capabilities. Informed by existing studies \cite{luo2024unlocking, Lei2024_RecExplainer}, we use LLMs to explain existing explainers' output and explore prompting with various information aspects. We build on prior LLM-as-a-judge paradigm findings \cite{zhang2024large, Li2025_alert, Zhao2025_adaptingLLMs} and present a large-scale analysis comparing user-friendly explanations generated with different information content.
We not only compare the LLM evaluations with human evaluators, but also provide analyses of rating behavior and guidelines for adopting the LLM-as-a-judge approach.

\section{Methodology}

In this section, we describe the methodology that we applied to generate and evaluate the explanations. We introduce the RS used to produce the recommendations and describe the procedure for generating and evaluating user-friendly explanations using an LLMs-as-a-judge framework and a user study.

\subsection{Recommendation Generation}

We select path-based recommendations and explanations as our use case since knowledge graphs ensure that recommendations and explanations are grounded in structured knowledge and provide information beyond user-item interactions. 
We selected the PGPR recommender~\cite{PGPR} with TransE embeddings~\cite{Bordes2013_TransE} to generate recommendations and explanations, an explainable path-based RS widely used as a baseline \cite{cafe, Park2022_reinforcement, Balloccu2022_postprocessing}. It leverages reinforcement learning to compute recommendations and outputs a list of (relation, entity type, entity) tuples as keys, along with a list of probability values for each hop in the path, representing the trace in the knowledge graph from the target user to the recommended item.

We selected the Mindreader dataset~\cite{Brams2020_mindreaderdata}, as it contains a rich knowledge graph for movie recommendations, with various node and edge types, including actors, directors, genres, movie topics, and the decade the movie was released. 
We performed a 60/20/20 split on the data and tuned the hyperparameters of the TransE embeddings and the PGPR recommender on the validation set using a random search with 50 trials.
The parameters for the TransE embeddings are: 
\begin{itemize*}
    \item[embedding dimension] 64,
    \item[lambda] 0.275,
    \item[learning rate] 0.089, and
    \item[negative samples]  15.
\end{itemize*}
For PGPR, we used:
\begin{itemize*}
    \item[gamma]  0.534,
    \item[learning rate]  0.043, and
    \item[max. iterations]  10000.
\end{itemize*}

We generate path-based recommendations and explanations for the top 10 recommended items of the test set users. 
The extracted paths from user to recommendation follow a pattern as in: \texttt{\{"[('self\_loop', 'user', 1455), ('watched', 'item', 'Interstellar'), ('has\_genre', 'category', 'Drama Film'), ('has\_genre', 'item', 'Dark Water')]": [1.0, 1.0, 2.0]\}}.
The extracted paths yield a Path Type Concentration (PTC) \cite{balloccu_reinforcement_2023} of 0.51, indicating a somewhat balanced path-type representation, but some types may be over- or underrepresented.

\subsection{User-friendly Explanation Generation}

To keep the LLM computation times manageable, we randomly sampled 100 users from the test set. Then, we extracted the publication year and a short synopsis for each recommended movie in this set (using its Wikidata ID and the TMDB API\footnote{\url{https://developer.themoviedb.org/docs/getting-started}}), to use them in the LLM prompts and user study instructions. As we could not match all IDs, we excluded 11 recommendations from our initial sample, leaving 989 user-recommendation pairs. 

To generate user-friendly explanations, we customize the prompts in a systematic way, along four information aspects, with the following dimensions and rationales:  
\begin{enumerate*}
\item \emph{background knowledge of the user} (no information, lay user, RS knowledge) to explore the LLM's ability to personalize explanations,
\item \emph{explanation goal} (yes, no) to explore whether LLMs can optimize explanations for a specific goal according to the mainstream evaluation procedure, and
\item \emph{user history} (yes, no) and
\item \emph{information about the RS and the path trace} (yes, no) to determine whether more information about these aspects yields better explanations, and to explore how explanations generated without any knowledge of these aspects are rated.
\end{enumerate*}
To keep the number of tokens in the prompts under control, we set a cutoff for the user history at 20 items, randomly sampled (c.f. \cite{Lei2024_RecExplainer}) and added the short movie summary extracted from TMDB.
The user history and the corresponding summaries were kept consistent across all prompt conditions.
Table \ref{tab:explanation_prompts} exemplifies these information aspects and their corresponding prompt block.
The task introduction and output-format instructions were the same across all dimensions. We combine the four information aspects into 18 prompt variations (c.f. Table \ref{tab:prompt-combos}) to generate different explanations.

We selected the model \emph{gpt-oss 120b}, a state-of-the-art open-weight LLM, to generate explanations from prompts with varying information content described above. We set the temperature to 0 to minimize variation in the results and to ensure reproducibility.

\begin{table*}[]
\caption{Overview of the information aspects and prompt blocks used to generate the different explanations. Introduction and output instructions were the same for all prompts.}
\label{tab:explanation_prompts}
\Description{Table with 3 columns and 6 rows. Each row represents a prompt element and the variations that were used to generate the explanations.}
\resizebox{0.99\linewidth}{!}{%
\begin{tabular}{@{}lp{2cm}p{17cm}@{}}
\toprule
\begin{tabular}[c]{@{}l@{}}Information\\ aspect\end{tabular} & Variations & Prompt block \\ \midrule
Introduction & - & Your task is to explain a movie recommendation to a user in human-readable text. The recommended movie is \emph{\{recommendation\}} \\ \midrule
User expertise & \begin{tabular}[c]{@{}l@{}}Lay user,\\RS knowl.,\\No info\end{tabular} & \begin{tabular}[c]{@{}p{17cm}@{}}Lay user: The user occasionally watches movies, but has no knowledge of how the RS works.\\ RS knowledge: The user has a general understanding of how a recommender system works and is familiar with basic machine learning~concepts.\end{tabular} \\ \midrule
User history & Yes, No & Here is a list of titles and short descriptions of movies the user has previously watched: \emph{\{user history\}}  \\ \midrule
RS info & Yes, No & The recommendations are generated by an algorithm traversing paths in a knowledge graph representing the previous interactions of the user with the items and the features of the items. Based on the interactions and item features, the recommender aims to find relations between a user and the items that the user has previously not interacted with and recommends the items with the highest probability of having such a relation. The recommender outputs a path in structured form, which is a list of four tuples: 1. The user that is receiving the recommendation and their ID in the dataset, 2. a movie the user has watched, 3. a feature of the movie or a user who also watched the same movie, 4. a recommended movie that shares the feature. The tuple of each element is structured as follows: (relation to the previous element, entity type, entity name). For this recommendation, the system returned this path: \emph{\{path-trace\}}  \\ \midrule
Expl. goal & Yes, No & The goal of the explanation is to increase the transparency of the RS for the user. \\ \midrule
\begin{tabular}[t]{@{}l@{}}Output\\instruction\end{tabular} & - & Write the explanation as if you would show it to a user. Only output the explanation; do not add any additional information. Be as detailed as possible, but keep the explanation short and concise with a maximum of one paragraph. \\
\bottomrule
\end{tabular}%
}
\end{table*}

\begin{table}[]
\caption{Overview of the different prompt conditions.
}
\label{tab:prompt-combos}
\Description{Table of six columns with 18 rows, each specifying the prompt elements that led to the 18 explanation conditions.}
\resizebox{\columnwidth}{!}{%
\begin{tabular}{@{}lllllp{4.2cm}@{}} %
\toprule
\begin{tabular}[c]{@{}l@{}}Condition\end{tabular} & User expertise & History & RS info & \begin{tabular}[c]{@{}l@{}}Expl. goal\end{tabular} & Label \\ \midrule %
0 & No & No & Path & No & Baseline  \\ %
1 & No & No & No & No  & No context \\ %
2 & Lay user & No & No & No & Lay \\ %
3 & RS knowl. & No & No & No & Knowl. \\ %
4 & No & Yes & Yes & No & Hist. + RS info \\ %
5 & Lay user & Yes & Yes & No & Lay + Hist. + RS info \\ %
6 & RS knowl. & Yes & Yes & No & Knowl. + Hist. + RS info  \\ %
7 & Lay user & Yes & No & No & Lay + Hist. \\ %
8 & RS knowl. & Yes & No & No & Knowl. + Hist. \\ %
9 & No & No & No & Yes & Transp. \\ %
10 & Lay user & No & No & Yes & Lay + Transp. \\ %
11 & Lay user & Yes & No & Yes & Lay + Hist. + Transp. \\ %
12 & RS knowl. & No & No & Yes & Knowl. + Transp. \\ %
13 & RS knowl. & Yes & No & Yes & Knowl. + Hist. + Transp. \\ %
14 & No & Yes & Yes & Yes & Hist. + RS info + Transp. \\ %
15 & Lay user & Yes & Yes & Yes & Lay + Hist. + RS info + Transp. \\ %
16 & RS knowl. & Yes & Yes & Yes & Knowl. + Hist. + RS info + Transp. \\ %
17 & No & Yes & No & No & Hist. \\\bottomrule %
\end{tabular}%
}
\end{table}

\subsection{LLMs-as-a-Judge Evaluation}
We first evaluated the 989 generated explanations using an LLMs-as-a-Judge framework. We used both open-weight and proprietary models and selected a variation of 14 models of different sizes. We ran every model with two temperature settings, 0.0 and 0.3. At the 0.3 temperature setting, we repeated each run five times to assess consistency in the evaluation. We selected the 0.0 setting to minimize output variance and the 0.3 setting to investigate whether the slight added variance would lead to more human-like behavior.

\paragraph{LLM Judges:} We used the following series of general-purpose, mixture of expert models (except for minimax, designed for coding):
\begin{description}[labelsep=0pt, leftmargin=0.2cm, topsep=0pt]
    \item[Large open-weight: ] mistral-large-3 (675b) \cite{mistral_large_3}, cogito-2.1 (671b) \cite{cogito_2_1}, deepseek-v3.1 (671b) \cite{deepseek_v3_1}, minimax-m2.5 (229b) \cite{minimax_m2_5}, gpt-oss (120b) \cite{gpt_oss}
    \item[Small open-weight: ] nemotron-3-nano (30b) \cite{nemotron_3_nano}, gemma3 (27b) \cite{gemma3}, phi4 (14b) \cite{phi4}, glm4 (9b) \cite{glm4}, llama3.1 (8b) \cite{llama3_1}, qwen3 (8b) \cite{qwen3}
    \item[Proprietary: ] claude-haiku-4-5 \cite{claude_haiku_4_5}, gemini-2.5-flash-lite \cite{gemini_2_5_flash_lite}, grok-4-1-fast-non-reasoning \cite{grok_4_1_fast}
\end{description}

\paragraph{Dependent variables:} Each model evaluated the explanations on six dimensions, namely \emph{satisfaction} (i.e., I am satisfied with the explanation.), \emph{scrutability} (i.e., The explanation would help me to identify when the recommendations are wrong.), \emph{transparency} (i.e., The explanation increases the transparency of the system.), \emph{perceived understanding} (i.e., The explanation helps me understand why I saw this recommendation.), \emph{persuasiveness} (i.e., The explanation convinces me to watch the recommended movie.), and \emph{recommendation quality} (i.e., The movie
recommendation is of high quality and fits my taste.). We selected these measures for the following reasons. Recent findings \cite{Balog2020_conflicting} indicate that among the established goals by \citet{tintarev2015explaining}, \emph{satisfaction}, \emph{scrutability}, and \emph{transparency} provide the most comprehensive evaluation of explanation quality.
Transparency is often measured by assessing if the explanation increases users understanding of why they see a recommendation, while the transparency goal definition states ``explain how the system works'' \cite{tintarev2015explaining}. Thus, we measure system transparency and the user's \emph{perceived understanding} of why an item was recommended separately.
Research has shown that LLM-generated explanations can persuade users to accept recommendations even when they are not factual \cite{maes2025mitigating}.
Thus, we purposefully added prompt conditions in which the explanation generation model cannot include truthful information about user preferences or the underlying recommendation approach. 
To test whether the lack of factuality affects the evaluator, we also assess participants' intention to follow the recommendation.
Lastly, we measure perceived recommendation quality, as it may affect the explanation evaluation.

All dependent variables were measured on a 5-point Likert scale, from 1 - ``not at all'' to 5 - ``fully agree''. 
LLMs were also asked to provide a reason for each rating. 
Each model was given the exact same information and instructions for formatting the output. The prompt is as follows, with the blue parts of the text replaced by user history, recommendation, or explanation:
\FrameSep3pt
\begin{framed}
    \textsf{Imagine it is Friday evening and you are looking for a movie to watch. These are some of the movies you have recently watched: \textcolor{blue}{\{3 movies from the user's history, each with a short synopsis\}}.
    You turn on your streaming service and see that it recommends you this movie along with the following explanation: \textcolor{blue}{\{title of recommended movie and explanation\}}.
    Rate how much you agree with the following statements on a scale from 1-5, 1 meaning 'not at all', 5 meaning 'fully agree'. Please also provide a short reason for your rating. Limit the reason to 1-2 sentences.
    1. The explanation increases the transparency of the system, 2. The explanation convinces me to watch the recommended movie, 3. The explanation helps me understand why I saw this recommendation, 4. I am satisfied with the explanation, 5. The explanation would help me to identify when the recommendations are wrong. 6. The movie recommendation is of high quality and fits my taste.
    Only provide the answer; do not add any additional text or explanations. Format the response in valid JSON in the form statement[n]: statement, rating[n]: rating, reason[n]: reason, where n is the number for each statement.}
\end{framed}

\subsection{User Evaluation}

We also evaluated a selection of the conditions in a user study with similar design. 
The goal of the user evaluation is to investigate how users' perceptions change when the explanations are generated with different information aspects in the generation prompts.

\emph{Independent variables:} We ran a 5 x 4 mixed study design, as follows. 
To avoid carryover or anchoring effects, we evaluate the prompt conditions, our independent variables, in a between-subjects design. While the LLMs evaluated all 989 explanations from all 18 prompt conditions, this was not feasible for the human evaluation.
To narrow down the number of explanations for the user study, we selected 5 prompt conditions and sampled 20 explanations from each condition based on their quality (cf. Section \ref{sec:explanationresults}) and LLM evaluation rating distributions (cf. Section 
\ref{sec:llmeval}). 
The selected prompt conditions are \emph{Baseline}, \emph{Hist.}, \emph{Lay + Transp.}, \emph{Lay + Hist.}, and \emph{Lay + Hist. + Transp.}.
The 20 explanations were sampled to cover high-, medium-, and low-variance explanations among the LLMs, and one group with the same randomly sampled recommendation across all conditions (five examples from each). These four types of samples were shown in a within-subjects design, each participant seeing two randomly chosen samples from each category to limit fatigue. 
\\\indent\emph{Dependent variables:} We used the six measures (i.e., \emph{satisfaction}, \emph{scrutability}, \emph{transparency}, \emph{perceived understanding}, \emph{persuasiveness}, and \emph{recommendation quality}) from LLMs-as-a-Judge evaluation, rated on a 5-point Likert scale from 1 (not at all) to 5 (fully agree). 

\emph{Study procedure:} Study participants received a brief description of the study and a consent form.\footnote{The user study was approved by the ethics committee of the institution.} Then, they answered a set of demographic questions and questions regarding their familiarity with movie RS. Next, they were shown eight explanations, one at a time. The order in which the explanations and the rating statements were shown was randomized. Participants were also asked to provide a reason for their rating in a single, optional input field.

\emph{Attention checks:} Attention checks were added (1) before the start of the evaluation, when collecting demographic information, and (2) as an additional rating item when evaluating each explanation.

\emph{Participants:} We recruited 234 participants via Prolific.\footnote{\url{https://www.prolific.com/}} The participants were required to be over the age of 18, be fluent in English, and have completed a minimum of 100 previous submissions with an approval rate of at least 98\%.
Because the reading ease scores of the generated explanations indicated that they might be difficult to comprehend, we also required participants to have at least a high school degree. After filtering out 18 participants who failed at least one attention check, we were left with 216 participants.

\section{Results}
\label{sec:results}
In this section, we present the results of our experiments.
We first analyze the generated explanations before discussing the LLM and human evaluation results.

\subsection{Explanation Analysis}
\label{sec:explanationresults}
We first performed a qualitative analysis of a subset of the generated explanations to understand to what extent the different prompt variations were incorporated by the generation model.
Then, we performed quantitative analyses of the generated explanations.

\emph{Qualitative analysis.}
%
To check whether the information blocks in the prompts were actually considered in the explanations, two authors independently evaluated a sample of 10 randomly selected user-recommendation pairs per condition, for a total of 180 pairs. For each condition, they evaluated the presence of the four possible information aspects in the explanation generation (i.e., recommendation information, user history, user experience, and explanation goal, c.f. Table \ref{tab:explanation_prompts}). According to Krippendorff's $\alpha$, the two annotators had perfect agreement in all conditions and aspects. 

In general, the qualitative analysis revealed that the generation model adapted the explanations according to the information in the prompt.
The lay-user explanations avoid technical jargon and highlight shared features between the user preferences and the recommended item.
Explanations targeted to users with RS knowledge often use more technical terms, such as algorithms, collaborative filtering, or knowledge graphs, while connecting user preferences to the recommendations.
Explanations targeting the transparency goal mention the connection between user preferences and the recommendation.
Prompt conditions in which the path-trace was present also led to explanations using the path information. 

The analysis, however, also revealed two types of limitations. First, the generation model systematically ignores user history information whenever the prompt also contains RS information.
The explanation contains the information from the path-trace, but no additional items from the provided user history.
However, when path-trace information is not provided, the explanation contains a subset of the movies from the user's history.
Second, when the prompt lacked information about the user's preferences, it would hallucinate a user history or a preference for a movie genre in the explanations. To ensure this behavior is not model-dependent, we generated explanations with the larger Cogito and the smaller Gemma models and observed the same pattern in both.

\emph{Quantitative analysis.}
We analyzed the generated explanations following \citet{munoz-ortiz_contrasting_2024} and report descriptive statistics, information on their lexical diversity, reading ease, and the similarity between explanations generated by the different prompts.
For lexical diversity, we report the Measure of Textual Lexical Diversity (MTLD) using lemmatized tokens, which is more robust to text length than the type-token ratio (TTR) \cite{munoz-ortiz_contrasting_2024}.

Unsurprisingly, the \emph{baseline} condition contains the shortest explanations with an average of 27.98 lemmas and 1.02 sentences. 
The longest explanations were produced by the prompt of the \emph{Knowl. + Hist. + Transp.} condition with an average of 149.76 lemmas and 4.18 sentences. The highest lexical diversity, with an MTLD score of 149.48, occurs in \emph{no context}, while \emph{Knowl. + Hist. + RS info + Transp.} is the least diverse with an MTLD of 69.16.

\begin{figure}
    \centering
    \includegraphics[width=0.99\columnwidth]{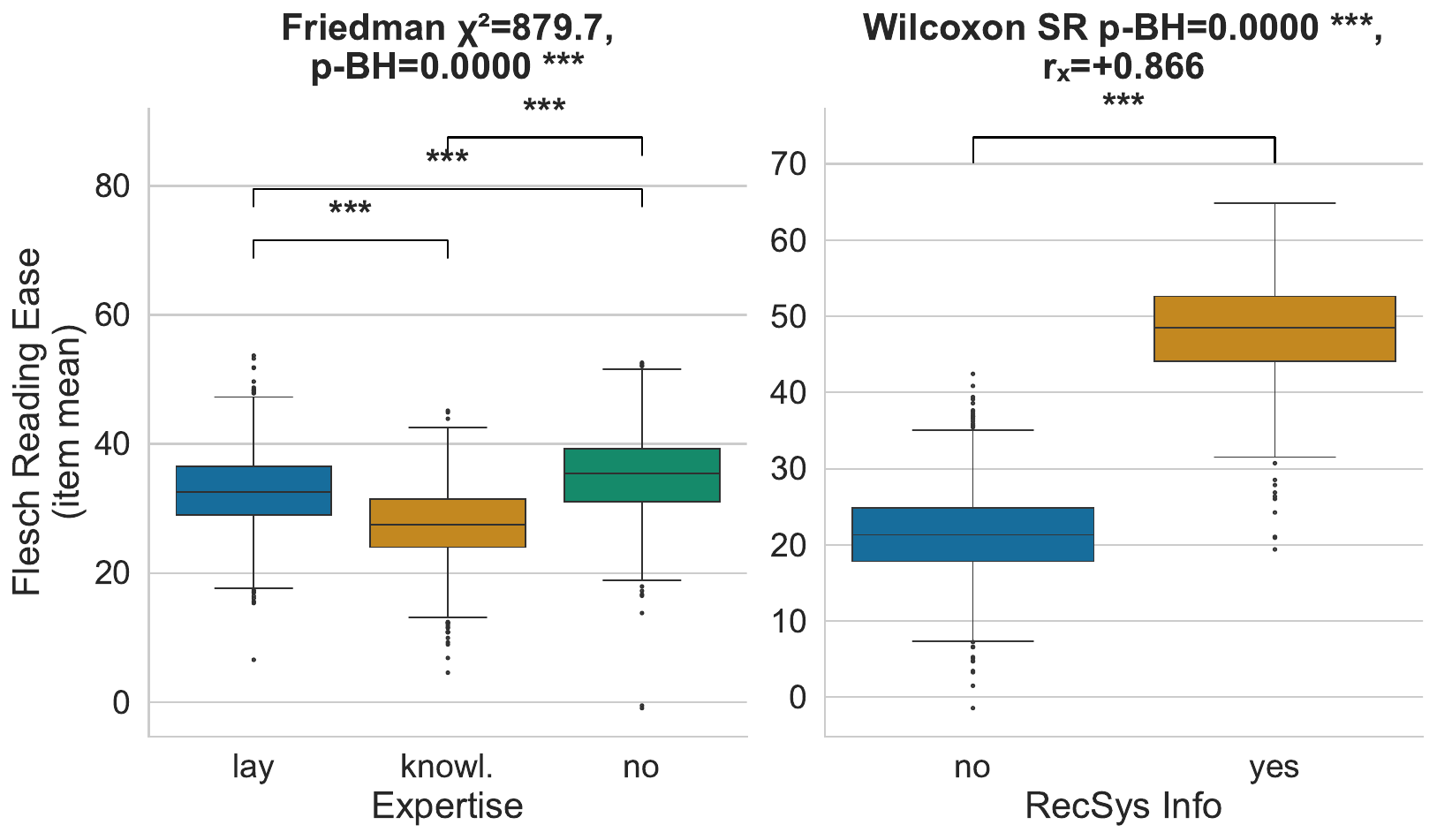}
    \caption{Reading ease comparison across conditions.}
    \Description{The figure shows two box plots. The left plot shows how the Flesh Reading Ease score changes across the three User Expertise variations. The box for the RS knowledge condition is the lowest, followed by the lay user condition, and the no info condition is the highest. The right plot shows the impact of RS info variations on the Reading Ease score: the box for the no-RS info variant is lowest.}
    \label{fig:reading-ease}
\end{figure}

We hypothesized that explanations generated for a target user with RS knowledge would yield a lower Flesh Reading Ease score than those targeting lay users.
The \emph{Knowl.} condition contains the lowest average Flesh Reading Ease score (16.38) among all conditions.
In Figure \ref{fig:reading-ease}, we compared the Reading Ease scores of the explanations for the variants of the different prompt blocks.
While the transparency goal does not impact the reading ease, all other prompt elements have a statistically significant effect.
We observed a greater impact on reading ease when adding information about the RS than when specifying the user's expertise.

\subsection{LLM-Based Explanation Evaluation}
\label{sec:llmeval}

Across all model outputs, we successfully parsed 8,944,644 ratings (99.7\%).
The remaining 0.3\% of the output was not parseable due to factors such as hallucinations, incomplete ratings, or an output structure different from the format specified in the prompt. These outputs came primarily from Deepseek due to hallucinations and from Nemotron due to differences in formatting and rating scales (e.g., it occasionally rated statements as \emph{true} or \emph{false}). 

\begin{figure}[ht]
    \centering
    \includegraphics[width=\columnwidth]{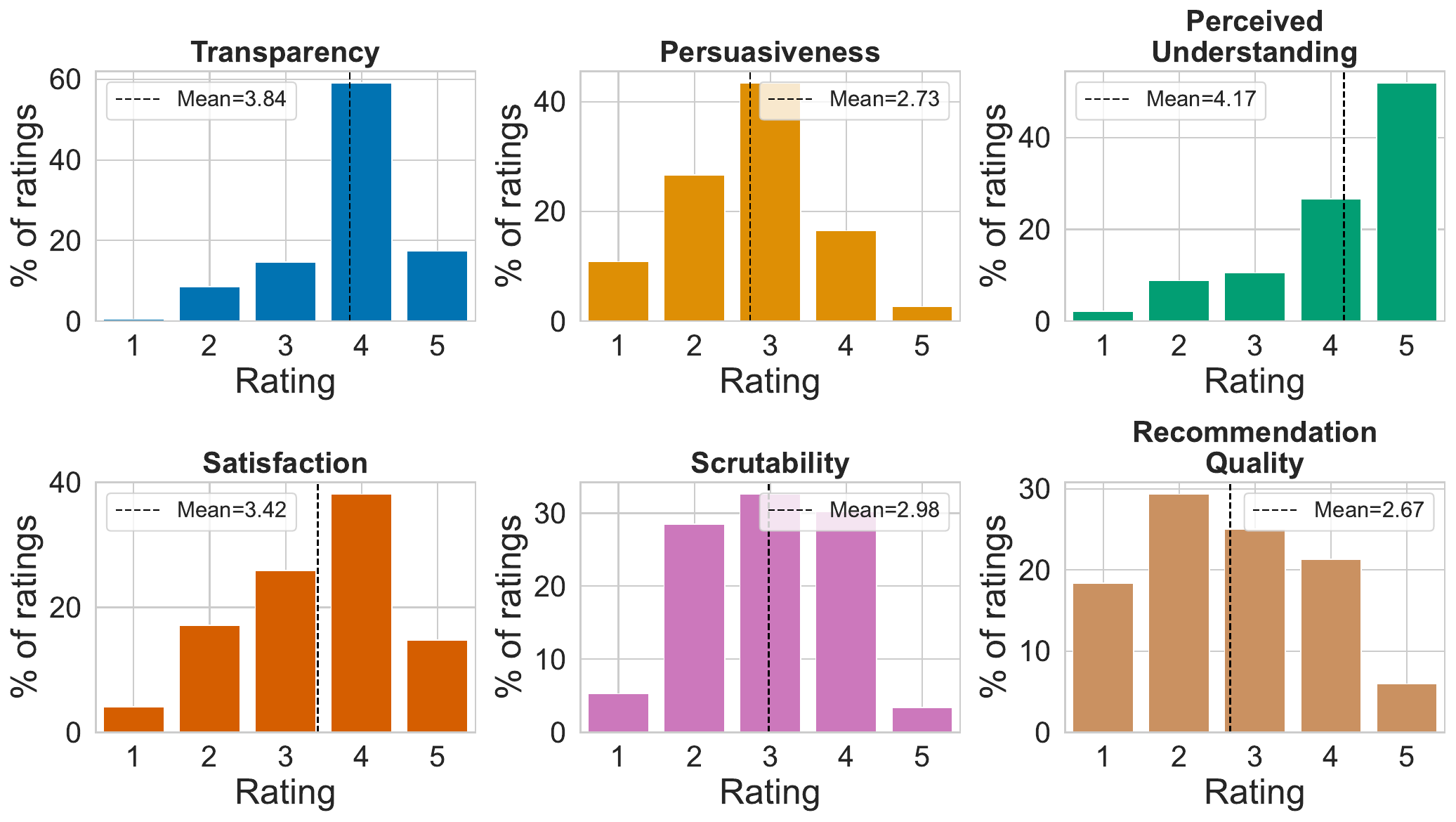}
    \caption{Rating distribution of the LLM evaluation.}
    \Description{The figure shows six bar charts, each depicting the rating distributions across one of the six evaluation constructs.}
    \label{fig:rating-distribution}
\end{figure}

Averaging the ratings across all models and runs (Figure~\ref{fig:rating-distribution}), we see that scrutability, persuasiveness, and recommendation quality were rated with means below 3, while satisfaction, transparency, and perceived understanding were rated more positively (means above 3). The perceived understanding stands out with a mean of 4.17. 

Looking at the rating distributions of the individual models, we see that \emph{some are more positive and others are more negative.}
According to the overall mean rating, Gemma is the most positive model, followed by Gemini, while Minimax and Claude are the most negative. Nevertheless, despite their differences in ratings, they still follow a similar pattern of rating distributions, which can be observed to varying degrees across all evaluation models.

We further observed that \emph{all models take explanations into account when rating recommendation quality.}
The evaluation prompt for each condition differed only in the explanation, yet each model rating of recommendation quality varied across conditions: Claude was the most stable, with a mean rating range of 1.74 to 2.55, while Deepseek showed the highest variance, with a mean of 1.33 to 3.11.

When comparing rating behavior across the two temperature settings, a Wilcoxon signed-rank test with Benjamini-Hochberg (BH) correction shows \emph{significant differences for all models, but the three proprietary ones.}
More precisely, a temperature setting of 0.3 in the Cogito model increased the average rating across all evaluation dimensions by 0.42, while the rating shift in all other proprietary models remains < 0.1.
Averaging the ratings across all models, we also find statistically significant effects on all evaluation dimensions except perceived understanding. Similar to the per-model comparison, the mean shifts only slightly by <0.03 points.

Regarding the consistency of repeated prompting ratings, we see \emph{substantial differences across models.}
Three LLMs (Gemma, Mistral, Phi) produce identical outputs in over 70\% of cases and have an overall mean standard deviation (SD) in ratings below 0.16.
Eight of our tested models, however, produce identical outputs in fewer than 50\% of cases, with Minimax being the most inconsistent (16.3\% identical outputs, 0.51 mean SD).

Due to inconsistent ratings when the temperature was set to 0.3 across the majority of models, we focus the analysis of inter-rater agreement (IRR) on ratings obtained at 0.0.\footnote{The IRR results at temp. = 0.3 only differ minimally from the reported results.}
We computed Krippendorff's $\alpha$ to evaluate the rating agreement among the LLMs.
Overall, the \emph{agreement is low.}
The value of 0.67, which is considered the lower bound for tentative conclusions, is never reached in any condition or evaluation aspect.
The highest agreement was observed for perceived recommendation quality, with values ranging from 0.19 for \emph{Transp.} condition to 0.3 for the \emph{baseline} and \emph{Hist. + RS info} conditions.
In the scrutability aspect, we observed disagreement among the models, with $\alpha$ values of 0.0 or lower.
Although their agreement is low, \emph{the mean Kendall-$\tau$-b across all 91 pairwise model combinations is 0.662, indicating a moderate correlation} (median 0.712, range [0.203, 0.869]).  
Due to the wide range of $\tau$ values, we examined the mean $\tau$ per model.
Cogito, Qwen, and Gemma have the highest mean rank correlations ($\tau$ = 0.743, 0.736, and 0.732, respectively).
The Claude and GLM models are at the other end of the spectrum, with mean $\tau$ values of 0.466 and 0.431, respectively.
We further examined the low rating agreement values and the inter-rater variance in the results.
First, we looked at the explanations that caused the most variance.
The 50 explanations with the highest mean SD are from 5 conditions, with \emph{Transp.} and \emph{Lay + Transp.} being the most prominent.
We further observe that the \emph{variance distribution varies for the evaluated aspects.}
While the variance distributions for persuasiveness and scrutability follow a bell curve, we find a bimodal distribution for transparency and perceived understanding.
We observe the same when examining the variance distributions across conditions.
Most curves are spread wide with some conditions (\emph{Hist. + RS info}, \emph{Lay + Hist. + RS info}, \emph{Hist. + RS info + Transp.}) following a bimodal distribution.

Next, we divided the models into small ($\leq 30B$ parameters) and large ($\geq 120B$ parameters), and treated the proprietary models as a separate category. We found that \emph{smaller models have a significantly higher mean rating} than the larger and proprietary ones (Kruskal-Wallis H = 29,078, p < 0.001; all pairwise Mann-Whitney U p < 0.001 after BH correction).
Furthermore, the \emph{large models have the highest mean Krippendorff's $\alpha$} (0.213) compared to the proprietary ($\alpha$=0.054) and small models ($\alpha$=0.025).

When analyzing the impact of the different information aspects in explanation generation prompts on ratings, we found that \emph{explanations generated with RS information generally received lower ratings, whereas those generated with user history received higher ratings.}\footnote{Even if history was not taken into account by the generation model (cf. Section \ref{sec:explanationresults}).}
Compared to the path-trace-only baseline, the explanations generated with RS information and the user history received, on average, 0.25 higher ratings.

\subsection{Human Explanation Evaluation}
\label{sec:human-eval}

The 216 participants were evenly distributed across conditions, with each condition ranging from 42 to 45 participants. 
29.2\% (63) were 35–44, 27.8\% (60), 25–34, 19.9\% (43), 45–54, 13.4\% (29), 55–64, 4.6\% (10), 18–24, 4.2\% (9), 65–74, and 0.9\% (2) above 75 years old.
117 participants identify as female, 95 as male, 3 non-binary, and 1 preferred not to disclose.
The majority (77.3\%) use movie streaming services at least weekly, and almost half (47.2\%) are moderately familiar with RS (19.5\% very familiar, 33.4\% slightly or not familiar). On average, each explanation was rated by 17.3 participants.

Looking at the overall rating distribution across all conditions and sampling strategies, persuasiveness was rated the lowest (mean = 3.13) and scrutability the highest (mean = 3.78).
Breaking down the results for the individual explanations, the condition \emph{Lay + Transp.} receives the highest overall mean rating across all six dimensions (M = 3.63), followed by \emph{Lay + Hist. + Transp.} (M = 3.57), \emph{Lay + Hist.} (M = 3.49), and \emph{Hist} (M = 3.49). The \emph{baseline} is consistently the lowest (M = 3.12).
Kruskal-Wallis tests confirm significant differences between the explanation conditions on all six evaluation dimensions (all H > 22, all p < 0.001) and on all ratings combined (H = 191.24, p < 0.001).

\begin{figure}[]
    \centering
    \includegraphics[width=\columnwidth]{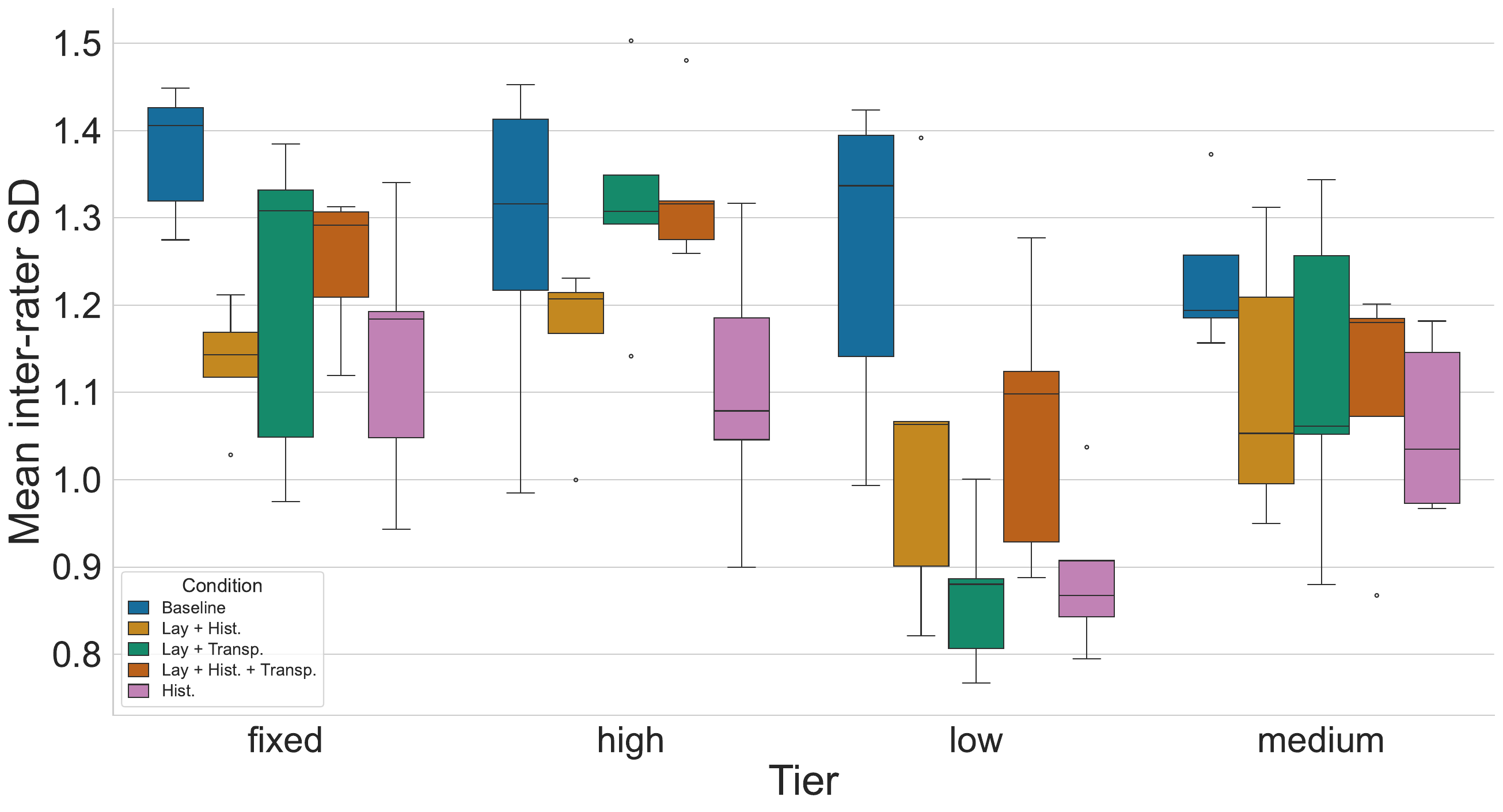}
    \caption{SD of IRR across conditions and sampling tiers.}
    \Description{The box plot shows four groups of boxes, one for each sampling strategy. Each group contains the distribution of IRR SD for the five evaluated conditions.}
    \label{fig:human-variance-condition-tier}
\end{figure}

Overall, similar to the LLM evaluators, \emph{participants show low IRR across conditions and evaluation dimensions.}
The highest agreement was among participants who saw the explanations of condition \emph{Lay + Transp} with $\alpha=0.111$. The lowest agreement ($\alpha=0.043$) was among participants who saw the \emph{Lay + Hist} explanations.
The persuasiveness and perceived recommendation quality ratings were slightly more in agreement, whereas the scrutability ratings showed the greatest disagreement.
When examining the IRR scores for each sampling tier, we find the highest disagreement ($\alpha=-0.006$) in the fixed user-item category and the high-variance sample ($\alpha=0.042$).
The ratings for the explanations sampled in the low-variance tier yield the highest agreement ($\alpha=0.079$), but the difference to the medium-variance tier is small ($\alpha=0.074$), and the overall difference across tiers is insignificant.
\emph{This indicates that human disagreement is not driven by selecting inherently ambiguous items: even explanations that LLM evaluators rated consistently still fail to produce meaningful agreement among human participants.}

In terms of ranking correlation, the pairwise Kendall's $\tau$ is significantly above zero in all conditions (bootstrap p < .001) but small, with a mean $\tau$ ranging from 0.077 (\emph{Lay + Hist.}) to 0.240 (\emph{Lay + Hist. + Transp.}).
The within-condition SD of pairwise $\tau$ is 0.67, reflecting a high variability across participant pairs: some pairs agree strongly while many show near-zero or negative rank correlation. 

Across all conditions, the mean SD of participants' ratings matches the LLM variance from which the explanation was sampled. 
When breaking down these results by condition (cf. Figure \ref{fig:human-variance-condition-tier}), we notice that the mean inter-rater SD is impacted by the condition.
The \emph{baseline} consistently has the highest mean SD. In the low-variance sampling tier, its mean is even higher than most conditions in the high-variance sampling tier.
For comparison, in the LLM evaluation, the \emph{baseline} explanations in the low-variance tier had a mean SD of 0.55 (1.26 for the user ratings), and those in the high-variance tier had a mean SD of 0.8 (1.28 for user ratings).

\subsection{Human vs. LLM Explanation Evaluation}

Our comparison of mean ratings and rankings between LLMs and humans (Table \ref{tab:rating_ranking_comp}) revealed the following. Overall, the ratings are similar, except for those for the \emph{baseline} condition, confirmed by the Wilcoxon signed-rank test (p=<0.0001, r=0.645).
Despite differences in the rating means, the baseline condition is the only one in which human and LLM evaluators agree on rank position.
While the human raters favored the \emph{Lay + Transp.} condition, in which user preferences and recommendation functionalities cannot be factually correct, the LLM raters preferred the explanations generated only based on the user history. This indicates that the generated explanations are plausible and convincing to humans. The differences in mean ratings among the first four ranks are small, though.

\begin{table}[]
\caption{Comparison of mean ratings and ranking of the explanations between human and LLM raters.}
\label{tab:rating_ranking_comp}
\Description{Table with six columns and five rows. Each row shows the human and LLM mean ratings, their difference, and the human and LLM rankings for one of the evaluated conditions.}
\resizebox{0.8\columnwidth}{!}{%
\begin{tabular}{@{}llllll@{}}
\toprule
Cond. & \begin{tabular}[c]{@{}l@{}}Human\\mean\end{tabular} & \begin{tabular}[c]{@{}l@{}}LLM\\mean\end{tabular} & Diff (H-L) & \begin{tabular}[c]{@{}l@{}}Human\\rank\end{tabular} & \begin{tabular}[c]{@{}l@{}}LLM\\rank\end{tabular} \\ \midrule
Lay + Transp.        & 3.629      & 3.548    & 0.080      & 1          & 2        \\
Lay + Hist. + Transp.        & 3.573      & 3.539    & 0.033      & 2          & 3        \\
Lay + Hist         & 3.495      & 3.474    & 0.021      & 3          & 4        \\
Hist.       & 3.494      & 3.584    & -0.090     & 4          & 1        \\
Baseline         & 3.118      & 2.608    & 0.510      & 5          & 5        \\ \bottomrule
\end{tabular}
}
\end{table}

\begin{figure}[]
    \centering
    \includegraphics[width=\columnwidth]{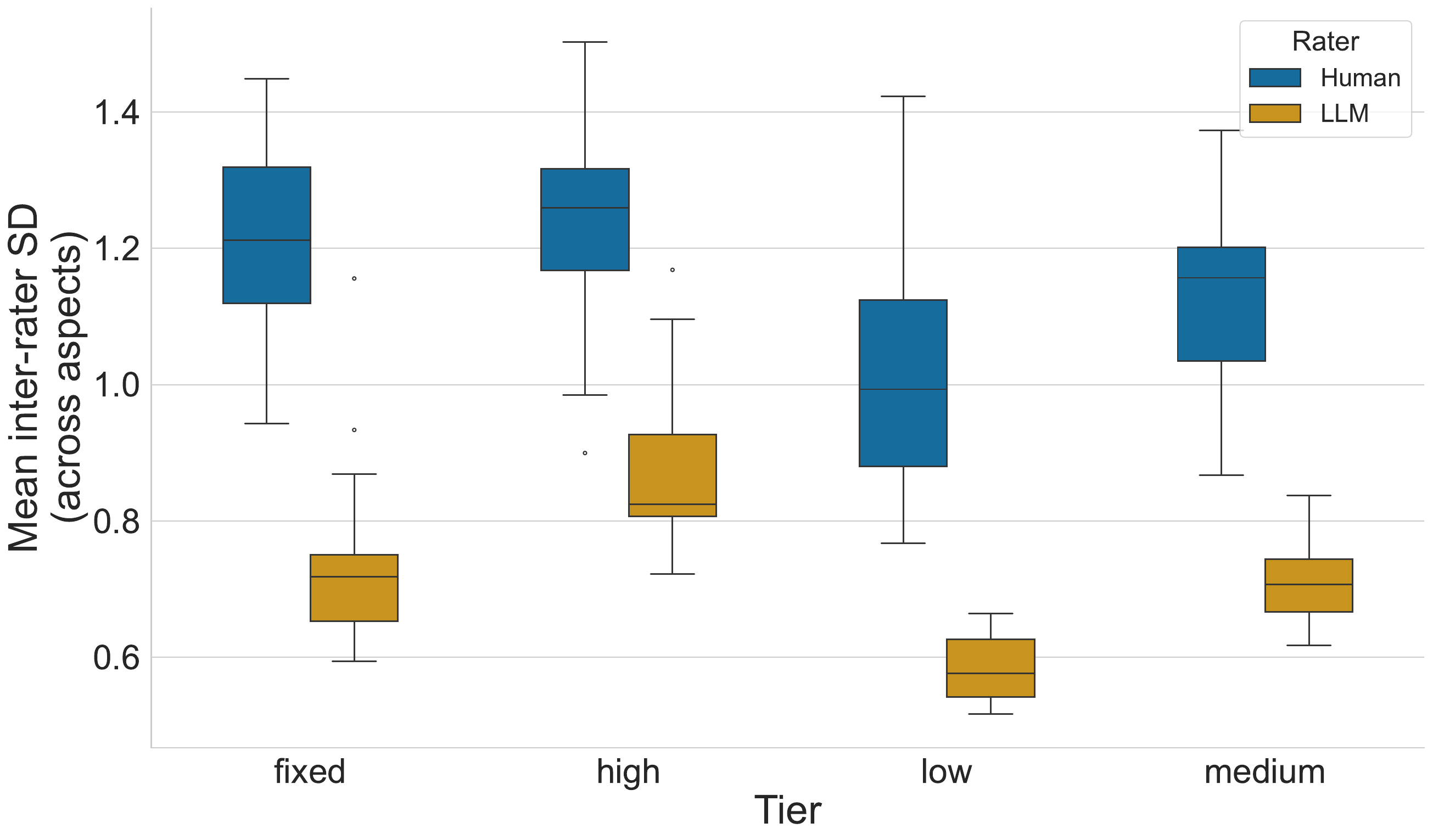}
    \caption{Human vs. LLM inter-rater SD for sampling tiers.}
    \Description{Boxplot comparing the human and LLM inter-rater standard deviation across the four sampling tiers. For each tier, the LLM has a significantly lower standard deviation.}
    \label{fig:human-llm-sd-comparison}
\end{figure}

In Figure \ref{fig:human-llm-sd-comparison}, we see that \emph{human raters disagree substantially more than LLM evaluators in every sampling tier.} 
A Mann-Whitney U test with BH correction across the sampling tiers shows that all four differences are significant (p < .001).

\begin{figure}[]
    \centering
    \includegraphics[width=0.7\columnwidth]{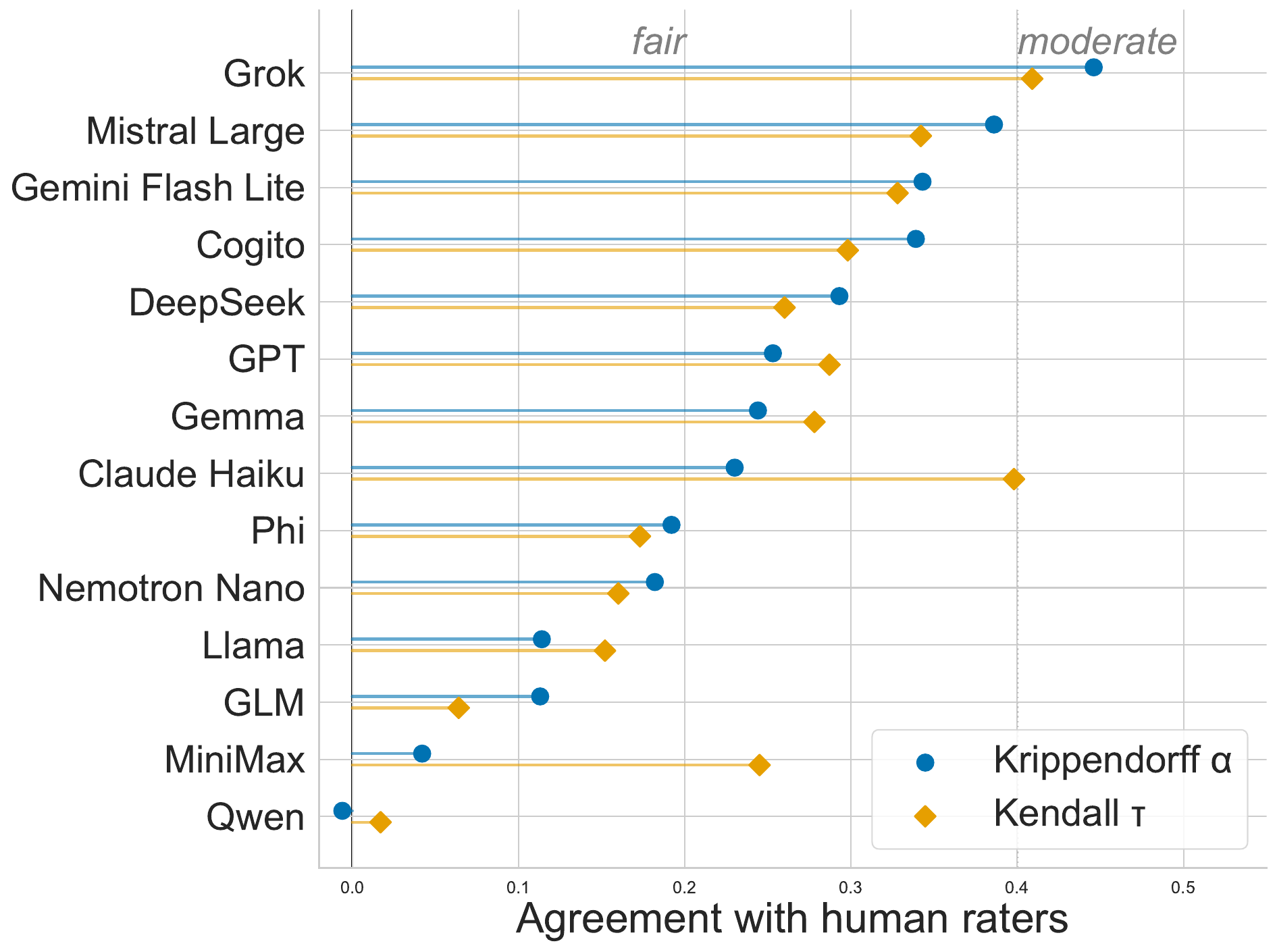}
    \caption{Rating agreement and ranking correlation between humans and each LLM evaluator.}
    \Description{Dotplot showing the Krippendorff alpha and Kendall tau values for each LLM when compared to the human ratings. Grok has the largest overlap, followed by Mistral. Qwen has the lowest alignment with the human ratings and rankings.}
    \label{fig:human-llm-agreement}
\end{figure}

Rating and ranking correlations between human and LLM evaluators reveal that \emph{larger models tend to align more closely with human raters.}
Figure \ref{fig:human-llm-agreement} shows that for both aspects, ranking correlation and rating agreement, Grok is the closest model to the human evaluation results (avg. Krippendorff $\alpha$ = 0.446, Kendall $\tau$ = 0.409).
Claude performs similarly in rank correlation but falls behind the Mistral model in rating agreement.

Regarding the evaluation dimension, \emph{scrutability shows the largest divergence in its correlation with other aspects}: the Human–LLM gap in pairwise Spearman $\rho$ is largest for Recommendation Quality $\leftrightarrow$ Scrutability ($\Delta\rho$ = +0.612), Persuasiveness $\leftrightarrow$ Scrutability ($\Delta\rho$ = +0.550), and Perceived Understanding $\leftrightarrow$ Scrutability ($\Delta\rho$ = +0.496). 
This indicates that humans perceive scrutability as tightly coupled with other quality dimensions, while LLMs treat it orthogonally. 

Finally, we examine the correlation between explanation properties and ratings.
The explanations' word count correlates positively with both human (Spearman $\rho$ = 0.458, p < .001) and LLM ratings ($\rho$ = 0.506, p < .001).
We also found a negative correlation of the ratings with the Flesch Reading Ease scores. 
Harder-to-read texts receive higher ratings (humans: $\rho$ = –0.307, p = .002; LLMs: $\rho$ = –0.533, p < .001). \emph{Both raters seem to prefer denser, detailed text.}
Humans and LLMs also rated the transparency of explanations higher when it was specified in the prompt, with a medium effect size (Mann-Whitney U, p-BH = 0.015 for both, r = -0.30).
For LLM raters, we found the same moderate effect of the goal prompt on perceived understanding ratings (Mann-Whitney U, p-BH = 0.015, r = -0.30), whereas there was no significant effect in the human ratings.

\section{Discussion and Recommendations}

From our results, we can extract key findings on explanation generation, LLM rating patterns, and their alignment with human raters, which we briefly discuss and derive recommendations from.

\emph{Explanation generation:} Our experiments with different information aspects in explanation generation blocks generally worked well. The qualitative results show adaptation to variations in user experience, and we found significantly higher transparency ratings when it was specified as a goal in the explanation-generation prompt.
The analysis also revealed challenges. When combining the user history and RS info prompt blocks, the LLM would not take the user history into account. Further investigation is needed to identify the cause of this behavior, but one possible reason is the length and information density of both prompt blocks.
Therefore, we recommend \emph{keeping the information in the prompt to a minimum} when combining several information elements.

\emph{LLM rating patterns:} Our results show similarities in the rating behavior of humans and LLMs. We found models following the same pattern, but some rate more positively and others more negatively. The analysis of perceived recommendation quality further indicates that LLMs are influenced by the information in the explanation, suggesting human-like behavior.
We did find differences depending on model size, though, with larger models showing better alignment with human raters.
While smaller models will likely improve, we currently recommend using \emph{models with $\geq$ 120B} parameters.

\emph{Alignment with humans:} We generally found a low IRR between humans and LLMs and a moderate ranking correlation.
Here, we want to discuss two factors that impact the agreement: the evaluation constructs and the explanations.
While we found relatively high agreement in the satisfaction and persuasiveness constructs, scrutability often generates disagreement.
One possible reason is that the evaluation setting, in which we asked evaluators to assess recommendations and explanations for other users, makes it difficult to determine when the system is wrong.
However, in this case, we would have seen a similar effect in the recommendation quality ratings.
We recommend \emph{carefully selecting and pre-testing the constructs that should be evaluated by LLMs.}
We also show that both humans and LLMs prefer explanation conditions that cannot contain factual information over the factual baseline and explanations containing RS info. 
This shows how plausible and convincing LLM-generated explanations could spread false information.
Non-factual information might be easier for users to spot when they receive explanations for their own recommendations, but we still recommend \emph{investigating methods to improve the user experience and acceptance of explanations generated by an explainable model.}
A starting point could be to ensure that the formatted explainers' outputs adhere to the properties of everyday human-to-human explanation, i.e., being contrastive, selective, social, and narrative-driven \cite{miller2019explanation}.

\section{Limitations and Future Work}
In this work, we evaluate explanations generated by a single model, using a single dataset, in a single domain.
While we do not think the quality of the recommendations affects our results, it should be noted that our results might not generalize to other domains or non-path-based explanation settings.
Future work should experiment with different settings to investigate how the results translate to other domains or explanation settings, such as counterfactuals or contrastive explanations.
We show that LLMs' ability to simulate user ratings also depends on the rating dimension. We sampled our rating constructs based on the literature, but future work should also explore other explanation goals and evaluation constructs. We further decided to sample a subset of the explanation conditions for the user study to keep evaluation costs manageable. We excluded, for example, explanations for users with RS knowledge.
As generative AI models continue to develop, future work should also explore LLMs' capabilities beyond the text modality.

\section{Conclusions}
In this paper, we present a large-scale experiment that systematically compares LLM- and human-based evaluations of natural-language recommendation explanations.
We first show how varying information content about the RS, the explanation goal, and the user can be used to generate explanations from a single explainable RS.
When evaluating these explanations with 14 LLMs across six rating dimensions, we found human-like rating patterns across the LLMs and moderate ranking correlations, but \emph{the overall rating agreement within and between LLMs is low}. 
A user study with 216 participants revealed that human ratings disagree significantly more than LLM ratings, but some models achieve moderate alignment with them in terms of IRR and rank correlation.
From our results, we derive a set of recommendations, namely (1) keep the information in explanation generation prompts minimal, (2) select larger models to align with human evaluators, (3) carefully select and test evaluation constructs, and (4) evaluate the factuality of the information in the explanations, as non-factual information might not be easy to spot for either humans or LLMs.
Following these guidelines, our findings suggest that some LLMs might produce a ranking of the explanations that is similar to human ranking. This would allow for the selection of a subset of explanations for human evaluation, \emph{making hybrid human-AI experimentation practical}.

\begin{acks}
    In preparing this paper, we used Grammarly for grammar and spelling correction and Claude for coding assistance. We acknowledge the contributions of the GenAI tools in enhancing our work and take full responsibility for the publication’s content.
We further thank the study participants for engaging with our experiment. 
\end{acks}

\bibliographystyle{ACM-Reference-Format}
\bibliography{lit}

\end{document}